\documentclass[useAMS,usenatbib,letters]{mn2e}
\usepackage[T1]{fontenc}
\usepackage[latin1]{inputenc}
\usepackage{graphicx}
\usepackage{amssymb}
\usepackage[fleqn]{amsmath}
\usepackage[T1]{fontenc}
\usepackage{pslatex}
\usepackage[totalwidth=480pt,totalheight=680pt]{geometry}

\makeatletter

\newcommand{\tdyn}{t_\mathrm{dyn}}
\newcommand{\tprec}{t_\mathrm{prec}}
\newcommand{\trlx}{t_\mathrm{rlx}}

\title[Fractals for Near-Keplerian Systems]{Fractal Geometry of Angular Momentum Evolution in Near-Keplerian Systems}
\author[M. A. G\"urkan]{M. Atakan G\"urkan\thanks{E-mail:
ato.gurkan@gmail.com}\\
Astronomical Institute ``Anton Pannekoek'',
University of Amsterdam, Kruislaan 403, 1098 SJ Amsterdam, The Netherlands\\
Leiden University, Leiden Observatory, P.O. Box 9513,
2300 RA Leiden, The Netherlands}
\begin{document}
\bibliographystyle{mn2e.bst} 

\date{Accepted 201x XXX xx. Received 201x XXX xx; in original form 201x XXX xx}

\pagerange{\pageref{firstpage}--\pageref{lastpage}} \pubyear{201x}

\maketitle

\label{firstpage}

\begin{abstract}
In this paper, we propose a method to study the nature of resonant
relaxation in near-Keplerian systems. Our technique is based on
measuring the fractal dimension
of the angular momentum trails and we use it to analyze the outcome of
$N$-body simulations. With our method, we can reliably determine the
timescale for resonant relaxation, as well as the rate of change of
angular momentum in this regime. We find that growth of angular momentum
is more rapid than random walk, but slower than linear growth. We also
determine the presence of long term correlations, arising from the
bounds on angular momentum growth. We develop a toy model that
reproduces all essential properties of angular momentum evolution.
\end{abstract}


\section{Introduction}\label{s:intro}

Dynamical systems where the gravitational potential is
dominated by a single mass are called {\em near-Keplerian}
\citep{2005ApJ...625..143T}, since the orbits of smaller masses
in such systems are very close to Keplerian conic sections.
A characteristic feature of these systems is the existence of
various distinct timescales over which the dynamical variables
change. For positions and velocities, this is the dynamical time
$\tdyn\sim(a^3/GM)^{1/2}$, where $M$ is the mass of the
central object and $a$ is a given orbit's semimajor axis.  The bound
orbits (ellipses) precess over the precession time $\tprec \sim (M/N
m_\star) \tdyn$, where $m_\star$ is the mass of smaller objects and $N$
is the number of these objects within $a$. The timescale for the evolution
of energy is the relaxation time, $\trlx \sim (M^2/m_\star^2 N \ln
\Lambda) t_{\rm dyn}$. These timescales form an hierarchy $\trlx
\gg \tprec \gg \tdyn$ and lead to three regimes for angular momentum
evolution \citep{1999imda.coll..391T}.

For short timescales ($t \lesssim \tdyn$), the changes in angular
momentum has no correlation since the torques felt at different parts
of the orbit vary rapidly. For intermediate timescales ($\tdyn
\lesssim t \lesssim \tprec$), where we are effectively averaging
over orbits, the changes in angular momentum are correlated, since
the configuration of the orbits change slowly. For long timescales
($t\gtrsim \tprec$), the correlation is lost again, since
the orbits are randomized as they precess in different
directions at different rates.
The presence of an intermediate regime with enhanced angular
momentum evolution was first recognized in the literature by
\cite{1996NewA....1..149R}. They called this process {\em
resonant relaxation}, since it is the result of a near resonance
between angular and radial frequencies of the orbit.  It plays
a central role in various scenarios that are proposed to take
place in the vicinity of supermassive black holes \citep[see,
e.g.,][]{1999MNRAS.309..447M,2008AIPC.1053...79A}. 

There are two major uncertainties regarding resonant relaxation. The
first one is the boundaries, especially the upper limit, for the
intermediate regime. The timescale over which the torques are
coherent is evidently related to the precession time $\tprec$,
but the detailed nature of this relationship is unknown. Order of
magnitude estimates for the timescales are not sufficient, since in
some cases the lifetimes of the stars in the systems under question
are comparable to the timescales of the dynamical processes \citep[for
an example at the Galactic centre, see][]{2009ApJ...697L..44M}.

The nature of evolution of angular momentum over intermediate
timescales is also uncertain. In this regime, since the
torques are correlated, the evolution of angular momentum
is not going to be {\em diffusive} ($\Delta L \propto \Delta
t^{1/2}$). \cite{1996NewA....1..149R} seem to suggest that the
evolution in this regime is {\em ballistic} ($\Delta L \propto \Delta
t$, see their Eq.6 and Fig.1) and this is also adopted by
\citet{2009ApJ...698..641E}; but their simulations contain too few
stars to draw conclusions in this regard. Indeed, in this work we find
that the nature of the angular momentum evolution in the intermediate
regime is between diffusive and ballistic.

We expose this behaviour and determine the timescales by carrying out
simplified simulations of near-Keplerian systems and a fractal analysis
of the evolution of angular momentum. In section \ref{sec:fractals},
we give a brief review of fractals and present a new method for
determining fractal dimension, suitable for trails obtained in
numerical simulations. In section \ref{sec:simulations} we describe
our simulations and the fractal analysis of the angular momentum
evolution. In section \ref{sec:toymodel}, we demonstrate that such
an evolution can be mimicked by a simple random walk that retains a
limited term memory. We discuss the limitations and implications of
our findings in section \ref{sec:discussion}. 

\section{Fractal Dimension}
\label{sec:fractals}
Fractals \citep{1982fgn..book.....M} are geometric objects that
exhibit a number of properties that make them suitable for modelling
physical processes and natural structures. The defining property of
fractals is that their Hausdorff-Besicovitch dimension (hereafter
fractal dimension, $D$) exceeds their topological dimension. A
number of methods for calculating the fractal dimension is given by
\cite{1982fgn..book.....M}, here we give a brief sketch and develop
a new one that is suitable for our purposes.

\begin{figure}
\includegraphics[width=\columnwidth]{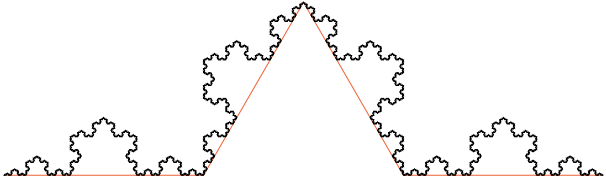}
\caption{The Koch curve, obtained by repeatedly transforming a
       line segment. The result of the first transformation is shown
       as a broken line in red, to demonstrate the self similar
       structure more clearly.  Each part of the Koch curve resting
       on the straight parts of this broken line is a smaller but
       an otherwise identical copy of itself.
       \label{fig:koch}
}
\end{figure}

A fractal is a self similar object; that is, part of it exhibits
similar properties to the whole, sometimes only in a statistical sense.
If a self similar object is made up of $n$ copies of itself, each
of which is smaller (in length) by $1/m$, then the object has fractal
dimension $D=\ln n/\ln m$. It is trivial to see that this leads to
correct numbers for self similar Euclidean objects. For example, any line
segment can be thought to be composed of $n=3$ identical copies of
itself, each of which is smaller by $1/3$ ($m=3$), giving $D=1$; or a rectangular
prism can be cut into $n=8$ identical copies of itself, each of which
is smaller (in length) by $1/2$ ($m=2$), giving $D=3$.
We can use this technique to calculate the dimension of a well known
fractal, the Koch curve (Fig. \ref{fig:koch}). This curve consists
of $n=4$ identical copies of itself\footnote{Koch curve can also
be seen as consisting of {\em two} copies of itself, each of which is
smaller by $1/\sqrt3$.}, each of which is smaller by $1/3$, leading
to a fractal dimension $D=\ln4/\ln3 \sim 1.262$.

Another way
to obtain the fractal dimension is to measure the length of a curve
in increasing detail. Since as we use smaller and smaller rulers, we will
be resolving more and more detail; the measured length of the curve $L$,
is a function of the ruler length $\varepsilon$ \citep{1967Sci...156..636M}:
\begin{equation}
L(\varepsilon) \propto \varepsilon^{1-D}\,.
\end{equation}
As an example, let us assume that we measure the length of the Koch
curve by a given ruler and obtain $L_0$. When we reduce the ruler
length by $1/3$, we shall be traversing the smaller copies in exactly
the same manner we traversed the whole curve. Since there are four
smaller copies, each of which will be measured to have a length
$1/3$ times the whole curve, we have $L'=4/3\times L$. To state in
more general terms, if we modify our ruler length by $\varepsilon'
\leftarrow (1/m) \varepsilon$, the measured length changes by
$L' \leftarrow (n/m) L$. It is easy to see that this scaling is
satisfied when $L \propto \varepsilon^{1-(\ln n/\ln m)}$. This
method of determining the (fractal) dimension of a curve readily
generalizes to real life curves, which are self similar only in a
statistical sense \citep{1967Sci...156..636M}.

For our purposes, it is more convenient to interpret a curve as the motion
trail of an object. As we check the position of the object on this trail
at decreasing time intervals, we will be resolving more
details and  the total trail length we calculate is
going to increase. In other words, the trail length is going to be a function
of the sampling interval $\chi$. For example, if we sample the
motion of an object on the Koch curve $4$ times, we will be measuring
the trail shown as the broken line in Figure \ref{fig:koch}, leading to
an increase in length by $4/3$, with respect to going from the beginning to
the end in one step. In more general terms, when the
sampling interval is modified by $\chi' \leftarrow (1/n)\chi$,
the measured length becomes $L' \leftarrow (n/m) L$. In this
approach, the fractal dimension is determined through the relation
\begin{equation}
L(\chi) \propto \chi^{\frac{1}{D}-1}\,. 
\end{equation}
This technique of measuring fractal dimension of a trail can be easily
applied to the results from dynamical simulations.

\section{Simulations}
\label{sec:simulations}
\subsection{Simulation Method and Parameters}

The system we simulated has three components. At the centre lies
a stationary supermassive black hole (SMBH) of mass $M_{\rm SMBH}
= 4\times10^6 M_{\sun}$. Around the SMBH, we have $N_{\rm field}
= 1200$ field stars each with mass $m_{\rm field} = 2.5 M_{\sun}$,
semi-major axes distributed in a powerlaw cusp $M_{\rm cusp}(r) \propto r^{3/2}$
from $a_{\rm min} = 0.0001$ to $a_{\rm max} = 0.03\,{\rm pc}$, with
eccentricities between $e_{\rm min} = 0$ and $e_{\rm max} = 0.95$,
distributed following the distribution function $g(e)\propto e$.
The final component is the massless test stars all with semimajor
axis $a_{\rm test} = 0.01\,{\rm pc}$, and eccentricities
$e_{\rm test} = 0.2$, $0.75$, and $0.85$ (12 stars each). Other
orbital elements (inclination, longitude of the ascending node,
argument of pericentre and mean anomaly) of all stars are picked at
random. The exact values chosen do not matter too much, since over the course
of the simulation the eccentricities are randomized.
We chose these parameters to have a system that somewhat resembles
the environment of S-stars observed at our Galactic centre. 
There are large uncertainties regarding the
star distributions at this environment \citep[][and
references therein]{2010ApJ...718..739M}, and 
$t_{\rm dyn}\ll t_{\rm prec} \ll t_{\rm rlx}$ hierarchy may not 
even exist there. However, this condition would hold for a region
around a SMBH that developed a  Bahcall-Wolf cusp
\citep{1976ApJ...209..214B,1977ApJ...216..883B}, so we expect
our method would be applicable to such systems.

The details of the code that is used for the simulations is going
to be explained in detail elsewhere, here we point out only the key
features. Both the field stars and the test stars feel the potential
of the SMBH, including the general relativistic (GR) correction that
leads to the prograde precession of the orbits\footnote{We use the
treatment of \citet[][their eq. 30]{1992AJ....104.1633S}}. The test
stars also feel the individual potential of the field stars, which
leads to retrograde precession of their orbits and changes in angular
momentum and energy. 
The field stars do not feel the individual
potential of each other, but instead see the potential of a smooth
cusp that is consistent with their distribution.  This approximation
decreases the time required for force computation significantly and
was already employed by \citet{1996NewA....1..149R} in their $N$-body
simulations. For the interactions between the test stars and field
stars we use a softening kernel ($\mathrm{K}_2$) of \citet{2001MNRAS.324..273D},
with a softening length $10^{-4}$\,pc. The units we adopted are
$G=M_\mathrm{field}=3000 M_{\sun} = 1\,\mathrm{pc} = 1$.

The extended mass distribution in the background cluster precesses the
orbits in a retrograde fashion, and the GR effects lead to prograde
precession. By our choice of parameters, these two effects cancel each
other for a star with $a\sim0.01\,\mathrm{pc}$ and $e\sim 0.6$. Mass
precession becomes more effective as semi-major axes get larger and
eccentricities get smaller, while the GR precession has the opposite
behaviour. 

We carry out the integration of the orbits using a high order
Runge-Kutta-Nystr\"om method, discovered by \citet[][their
SRKN$^b_{11}$]{2000JCAM..142..313B}. We split the Hamiltonian
into Keplerian and perturbation parts \citep{1991CeMDA..50...59K}
to increase efficiency and avoid spurious precession. This scheme
advances the Keplerian orbital elements correctly, except for the
truncation and the roundoff errors (the largest accumulated error
per star amounts to $\sim 10^{-5}$ over the course of the whole
simulation). In particular, the evolution of the Runge-Lenz vector does
not exhibit a linear drift as in the case of potential energy-kinetic
energy splitting \citep{1995ApJ...443L..93H}. Our treatment of the GR
perturbation leads to a small error in mean motion, but since we are
interested in changes that take place over many orbits, this error is
not important.  We use shared adaptive timesteps, but time-symmetrize
the integration with the method of \citet{1995ApJ...443L..93H}. Our
simulations last 30 code units which corresponds to a few precession
times of the slowest precessing test stars.

\begin{figure}
\includegraphics[width=\columnwidth]{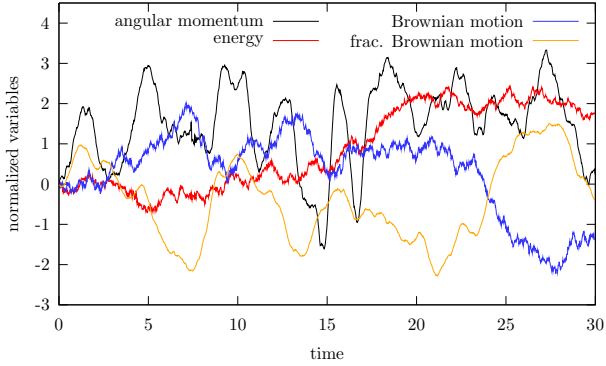}
\caption{The evolution of angular momentum (black) and energy (red)
	for a test star with initial eccentricity $e_0=0.75$. For
	comparison, a Brownian motion (blue) and a fractional Brownian
	motion (yellow; $c=18$, $n=4000$) data as described in
	sec. \ref{sec:toymodel} are also plotted. All curves are
	sampled at the same abscissae, and the ordinates are adjusted
	to have unit variance.
\label{fig:E_and_J}}
\end{figure}
We record the energy and angular momentum of each test star throughout
the simulation. In Figure \ref{fig:E_and_J} we show the evolution of
these quantities for a star with initial eccentricity $e_0=0.75$. Even
by eye, it is possible to tell that these quantities show a different
behaviour. For comparison, in this figure we also plot the curves
generated by Brownian and fractional Brownian motion, as described
in Section \ref{sec:toymodel}.

\subsection{Analysis of Simulation Results}
\begin{figure}
\includegraphics[width=\columnwidth]{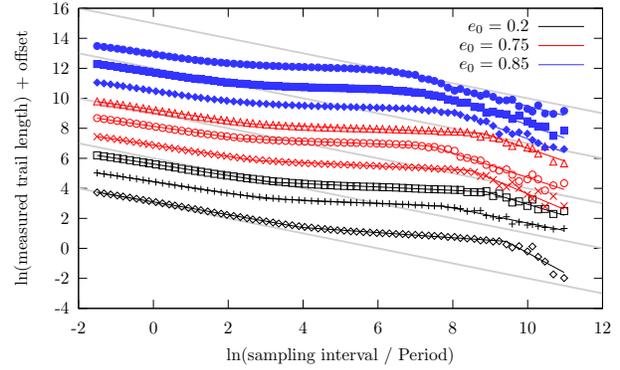}
\caption{Measured trail lengths as a function of duration of the
	sampling intervals for angular momentum trails and the
	twice-broken power-law fits made to them.
	Sets are artificially offset from each other to avoid confusion.
	The gray lines in the background are to lead the eye and have
	slopes $-1/2$.
       \label{fig:trail_data} }
\end{figure}
We measure the length of the angular momentum trail of each test star
by sampling it at intervals ranging from the full length of the
simulation down to a fraction of the orbital period. The dependence of 
the total measured trail length on the sampling interval is shown in
Figure \ref{fig:trail_data} for a few stars with different initial
eccentricities. Starting from long sampling intervals and moving towards
shorter ones, a number of features can be observed on this figure:
\begin{itemize}
\item For long timescales, the curves do not have the slope $-1/2$
(corresponding to dimension  $D=2$, the value for Gaussian random walks
\citep{1982fgn..book.....M}), but are somewhat steeper. This is to be
expected since the energy of an orbit changes through relaxation and
this process is much slower; hence the angular momentum evolution is
bounded unlike a true random walk, and {\em cannot be described by
simple diffusion}, even for long timescales.
\item There is a marked transition to a more coherent motion as
indicated by a decrease in the slope. The point of this transition can
be determined with reasonable accuracy for a given star, but it is not
common for all stars. 
\item Even though the slope decreases, it never becomes zero; hence
{\em the evolution of angular momentum is never ballistic
($\Delta L \not\propto \Delta t$)}.
\item This slope can also be determined with reasonable accuracy for a
given star, but varies from star to star.
\item Looking at the eccentricity evolution of the stars reveals that
the transition point is later for the stars that moved into $e\sim
0.6$ region where precession is slow. This is in harmony with the expectation
that for more slowly precessing stars, the coherent torques last longer.
\item The slope increases again for short timescales, but much before
the period of the stars is reached. This randomization of the torques is
a result of the stochastic nature of the processes that develop the
torques and dominates the shorter timescale randomization that would result
from orbital motion, at least down to the timescales we resolve.
\end{itemize}

Our results verify the presence of an intermediate regime, where
the angular momentum evolution is enhanced. The evolution of angular
momentum in this regime is not as rapid as ballistic growth $\Delta
L \propto \Delta t$, but more rapid than diffusive growth $\Delta
L \propto \Delta t^{1/2}$. This manner of evolution can be seen as
the generalization of Brownian motion called {\em fractional
Brownian motion}.  \cite{1968SIAMRev.10..422M} describes various
properties of this motion, along with applications. \cite{1982fgn..book.....M}
gives further generalizations and methods to produce such curves. Even
though these approaches are mathematically elegant and complete,
in the next section we propose a simpler model that is easier to
attach a physical interpretation.

\section{A Simple Toy Model for Angular Momentum Evolution}
\label{sec:toymodel}
One dimensional Brownian motion for a particle's position $P(t)$ can be
generated as follows.  Let the motion over $\Delta t$ consist of $N$
steps with equal duration $\delta t=\Delta t/N$. We choose an initial
value $P_0$ and at each step either increase or decrease the value
of $P$ by $\delta P$. We decide which action to take by generating a
sequence of random numbers $X_i$ uniformly sampled from the interval
$[-0.5, 0.5]$ and choosing a threshold value $q=0$. At each step,
if $X_i<q$ we increase the value of $P$ and otherwise decrease it.
To make the variance of the motion independent of the number of
steps we choose $\delta P \propto 1/\sqrt{N}$.  This scheme leads
to Brownian motion (Gaussian random walk) for small $\delta t$, i.e,
a large number of steps \citep[][Chap. 16]{2003falconer..book}. It
can be extended to a vector variable by letting each component perform
independent Brownian motions.

We can introduce correlations between the increments of the variable
$P$ by using a ``repository''. For this, we keep track of the last $n$
values of $X_i$ and let our threshold value be proportional to their
average $q=c\left<X_i\right>_n$, where $c$ is some constant. Here,
$n$ determines the length of the correlations and $c$ determines
their strength. We generated a few sets of data this way and measured
the lengths of the resulting trails with differing sampling durations.
\begin{figure}
\includegraphics[width=\columnwidth]{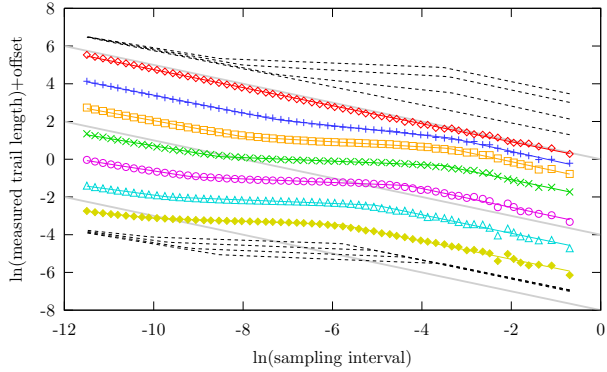}
\caption{Measured trail lengths as a function of duration of the
	sampling intervals for various generated random walk data
	sets and the twice-broken power-law fits made to them. The
	set in the middle (green) has $c=18$, $n=4000$. The sets
	above it have varying $c=12,6,0$ and the sets below it have
	varying $n=2000, 1000, 500$ (see the text for explanation
	of these parameters). All sets have $N=200\,000$ points. To
	avoid confusion, sets are artificially offset from each other;
	but to make comparison easier, top four and bottom four sets
	are redrawn with dashed black lines at the top and bottom of
	the figure with identical offsets.
	The gray lines in the background are to lead the eye and have
	slopes $-1/2$.
       \label{fig:param_data} }
\end{figure}

The curves generated this way (Fig.~\ref{fig:param_data}) show
very similar characteristics to angular momentum evolution. They
exhibit an intermediate regime with lowered slope, whereas for long and
short timescales the motion is more randomized. The upper bound of the
intermediate regime is determined by the parameter $n$: the break occurs
when the sampling interval matches $n\times\delta t$. The other
parameter $c$ determines the slope in the intermediate and short
timescale regime. Larger $c$ leads to a more coherent motion and a slope
closer to $0$.

\begin{figure}
\includegraphics[width=\columnwidth]{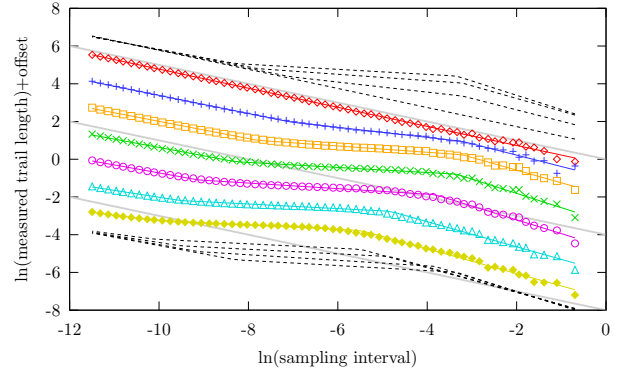}
\caption{Same as Fig. \ref{fig:param_data} but for bounded random
	walks. The slopes for long sampling intervals are steeper
	than $-1/2$.
       \label{fig:param_data_bounded} }
\end{figure}
We also generated similar data with bounded random walks. For
those, we started with a (vector) variable of unit magnitude
$|\mathbf{P}(t=0)|=1$, and whenever a step led to $|\mathbf{P}|>1.5$,
we took that step in the opposite direction.  This limit roughly
corresponds to the limit experienced by an orbit starting with
initial eccentricity $e_0=0.75$, $J_\mathrm{circular}/J_0 \sim
1.5$. The results from this bounded random walks are shown in Figure
\ref{fig:param_data_bounded}, exhibiting the long term correlations
similar to angular momentum evolution.

\section{Discussion}\label{sec:discussion}
In this work, we studied the evolution of angular momentum in near-Keplerian
systems by analyzing the outcome of $N$-body simulations. The
simulation method we use incorporates certain approximations.
Our test stars are massless, so they do not cause a back-reaction
in the surrounding cluster. Furthermore, the field stars all have
the same mass. A mass spectrum would change the granularity of the
background potential, affecting the applied torques and possibly
the coherence timescale. Finally, since the field stars do not see
the granularity of their own potential, their angular momenta do not
evolve. This decreases the rate of randomization of the background
cluster, since vector resonant relaxation can change the orientation
of the orbits on timescales comparable to mass and GR precession
for some systems\footnote{We thank an anonymous referee for pointing
out this shortcoming.} (for a comparison of these timescales for the
Galactic centre, see \citet{2010arXiv1006.0001K}). If the torques on
a star mainly change because of the rearrangement of the background
cluster, rather than the reorientation of the orbit, this decrease in
randomization would alter the rate of angular momentum evolution. All
these approximations limit the domain of applicability of our $N$-body
simulation approach; however, none of them alter the essential
mechanism by which the angular momentum evolves.

We analyzed this evolution by calculating the fractal
dimension of the angular momentum trail.  With this method it
is possible to reliably determine the onset of and the rate of
evolution in different regimes. A key result of our analysis
is that the evolution of angular momentum is neither diffusive
nor ballistic in any regime, as was previously assumed in other
studies \citep[e.g.][]{2006ApJ...645.1152H,2009ApJ...698..641E}.
This seems to contradict the results of the numerical experiments
done by \cite{1996NewA....1..149R}.  The reason for this discrepancy
is not clear, but we speculate that it arises from the low number of
stars used in that work.

We also developed a toy model that reproduces the features of
angular momentum evolution. This model has adjustable parameters with
clear physical interpretations. The relation between the appropriate
values of these parameters and the physical variables requires more
detailed and extensive analysis, which is currently underway. Apart
from studying different initial conditions, we also plan to analyze
the components of the torque parallel and perpendicular to the angular
momentum separately. The nature of these torques can be very different
\citep{1996NewA....1..149R,2007MNRAS.379.1083G}, so a separate analysis
should lead to a better understanding.

\citet{2002HSJ...47.573K} developed fractional Gaussian noise
generators (FGNGs) similar to our repository model. In that work
he compares autoregressive moving average (ARMA) models to FGNGs,
and finds that ARMA models are inferior for describing long-term
correlations.  These models need to be modified to have long term
memory \citep{2000tsaa..book.....S}, to describe angular momentum
evolution in near-Keplerian systems. Alternatively, they can be
used to describe short term correlations for torques, and long
term correlations can be introduced by taking physical bounds
into account, as is done here. After this paper was submitted,
\citet{2010arXiv1010.1535M} also submitted a paper containing their
analysis of this problem with ARMA models. They use ARMA(1,1) model,
which fixes the value of the autocorrelation function for very large
timescales to zero (see their figure 4) and hence does not lead to
any correlations beyond a given time.

The computer programs used to generate and analyze the data are
available from the author upon request.

\section*{Acknowledgments}

This work is supported by a Netherlands Organization for Scientific
Research (NWO) Veni Fellowship. Most of the simulations were done on
Lisa cluster, maintained by SARA, the Dutch National High Performance
Computing and e-Science Support Center. I thank all SARA staff
for doing an exceptional job for maintaining this cluster and in
particular to Walter Lioen for his help. Usage of Lisa was possible
through a grant (client number 10450) to Simon Portegies Zwart.
I am grateful to Clovis Hopman, Yuri Levin, Ann-Marie Madigan and
\.Inan\c{c} Adagideli for fruitful discussions on this topic, and an
anonymous referee for comments that improved this paper.

\bibliography{frac_pap}

\begin{thebibliography}{}

\bibitem[\protect\citeauthoryear{{Alexander}}{{Alexander}}{2008}]{2008AIPC.105%
3...79A}
{Alexander} T.,  2008, in {S.~K.~Chakrabarti \& A.~S.~Majumdar} ed., AIP
  Conference Series Vol.~1053, {The Galactic Center as a laboratory for extreme
  mass ratio gravitational wave source dynamics}.
p.~79

\bibitem[\protect\citeauthoryear{{Bahcall} \& {Wolf}}{{Bahcall} \&
  {Wolf}}{1976}]{1976ApJ...209..214B}
{Bahcall} J.~N.,  {Wolf} R.~A.,  1976, \apj, 209, 214

\bibitem[\protect\citeauthoryear{{Bahcall} \& {Wolf}}{{Bahcall} \&
  {Wolf}}{1977}]{1977ApJ...216..883B}
{Bahcall} J.~N.,  {Wolf} R.~A.,  1977, \apj, 216, 883

\bibitem[\protect\citeauthoryear{{Blanes} \& {Moan}}{{Blanes} \&
  {Moan}}{2000}]{2000JCAM..142..313B}
{Blanes} S.,  {Moan} P.~C.,  2000, Journal of Computational and Applied
  Mathematics, 142, 313

\bibitem[\protect\citeauthoryear{{Dehnen}}{{Dehnen}}{2001}]{2001MNRAS.324..273%
D}
{Dehnen} W.,  2001, \mnras, 324, 273

\bibitem[\protect\citeauthoryear{{Eilon}, {Kupi} \& {Alexander}}{{Eilon}
  et~al.}{2009}]{2009ApJ...698..641E}
{Eilon} E.,  {Kupi} G.,    {Alexander} T.,  2009, \apj, 698, 641

\bibitem[\protect\citeauthoryear{{Falconer}}{{Falconer}}{2003}]{2003falconer..%
book}
{Falconer} K.~J.,  2003, {Fractal Geometry: Mathematical Foundations and
  Applications}.
John Wiley \& Sons, West Sussex

\bibitem[\protect\citeauthoryear{{G{\"u}rkan} \& {Hopman}}{{G{\"u}rkan} \&
  {Hopman}}{2007}]{2007MNRAS.379.1083G}
{G{\"u}rkan} M.~A.,  {Hopman} C.,  2007, \mnras, 379, 1083

\bibitem[\protect\citeauthoryear{{Hopman} \& {Alexander}}{{Hopman} \&
  {Alexander}}{2006}]{2006ApJ...645.1152H}
{Hopman} C.,  {Alexander} T.,  2006, \apj, 645, 1152

\bibitem[\protect\citeauthoryear{{Hut}, {Makino} \& {McMillan}}{{Hut}
  et~al.}{1995}]{1995ApJ...443L..93H}
{Hut} P.,  {Makino} J.,    {McMillan} S.,  1995, \apjl, 443, L93

\bibitem[\protect\citeauthoryear{{Kinoshita}, {Yoshida} \& {Nakai}}{{Kinoshita}
  et~al.}{1991}]{1991CeMDA..50...59K}
{Kinoshita} H.,  {Yoshida} H.,    {Nakai} H.,  1991, Celestial Mechanics and
  Dynamical Astronomy, 50, 59

\bibitem[\protect\citeauthoryear{{Kocsis} \& {Tremaine}}{{Kocsis} \&
  {Tremaine}}{2010}]{2010arXiv1006.0001K}
{Kocsis} B.,  {Tremaine} S.,  2010, arXiv astro-ph, 1006.0001

\bibitem[\protect\citeauthoryear{{Koutsoyiannis}}{{Koutsoyiannis}}{2002}]{2002%
HSJ...47.573K}
{Koutsoyiannis} D.,  2002, Hydrological Sciences Journal, 47, 573

\bibitem[\protect\citeauthoryear{{Madigan}, {Hopman} \& {Levin}}{{Madigan}
  et~al.}{2010}]{2010arXiv1010.1535M}
{Madigan} A.,  {Hopman} C.,    {Levin} Y.,  2010, arXiv astro-ph,1010.1535

\bibitem[\protect\citeauthoryear{{Madigan}, {Levin} \& {Hopman}}{{Madigan}
  et~al.}{2009}]{2009ApJ...697L..44M}
{Madigan} A.,  {Levin} Y.,    {Hopman} C.,  2009, \apjl, 697, L44

\bibitem[\protect\citeauthoryear{{Magorrian} \& {Tremaine}}{{Magorrian} \&
  {Tremaine}}{1999}]{1999MNRAS.309..447M}
{Magorrian} J.,  {Tremaine} S.,  1999, \mnras, 309, 447

\bibitem[\protect\citeauthoryear{{Mandelbrot}}{{Mandelbrot}}{1967}]{1967Sci...%
156..636M}
{Mandelbrot} B.~B.,  1967, Science, 156, 636

\bibitem[\protect\citeauthoryear{{Mandelbrot}}{{Mandelbrot}}{1982}]{1982fgn..b%
ook.....M}
{Mandelbrot} B.~B.,  1982, {The Fractal Geometry of Nature}.
Freeman, San Francisco

\bibitem[\protect\citeauthoryear{{Mandelbrot} \& {van Ness}}{{Mandelbrot} \&
  {van Ness}}{1968}]{1968SIAMRev.10..422M}
{Mandelbrot} B.~B.,  {van Ness} J.~W.,  1968, SIAM Review, 10, 422

\bibitem[\protect\citeauthoryear{{Merritt}}{{Merritt}}{2010}]{2010ApJ...718..7%
39M}
{Merritt} D.,  2010, \apj, 718, 739

\bibitem[\protect\citeauthoryear{{Rauch} \& {Tremaine}}{{Rauch} \&
  {Tremaine}}{1996}]{1996NewA....1..149R}
{Rauch} K.~P.,  {Tremaine} S.,  1996, New Astronomy, 1, 149

\bibitem[\protect\citeauthoryear{{Saha} \& {Tremaine}}{{Saha} \&
  {Tremaine}}{1992}]{1992AJ....104.1633S}
{Saha} P.,  {Tremaine} S.,  1992, \aj, 104, 1633

\bibitem[\protect\citeauthoryear{{Shumway} \& {Stoffer}}{{Shumway} \&
  {Stoffer}}{2000}]{2000tsaa..book.....S}
{Shumway} R.~H.,  {Stoffer} D.~S.,  2000, {Time Series Analysis and Its
  Applications}.
Springer-Verlag, New York

\bibitem[\protect\citeauthoryear{{Tremaine}}{{Tremaine}}{1999}]{1999imda.coll.%
.391T}
{Tremaine} S.,  1999, in Impact of Modern Dynamics in Astronomy, IAU Colloquium
  172, {Resonant relaxation}.
p.~391

\bibitem[\protect\citeauthoryear{{Tremaine}}{{Tremaine}}{2005}]{2005ApJ...625.%
.143T}
{Tremaine} S.,  2005, \apj, 625, 143

\end{thebibliography}

\bsp

\label{lastpage}
\end{document}